\documentclass{aa}
\usepackage{graphicx}
\usepackage{natbib}
\usepackage{amsmath,amssymb}

\bibpunct{(}{)}{;}{a}{}{,} % to follow the A&A style

\newcommand{\rhod}{\ensuremath{\rho_{\mathrm{d}}}}
\newcommand{\fdrag}{\ensuremath{f_{\mathrm{drag}}}}
\newcommand{\ssD}{\ensuremath{_{\mathrm{D}}}}
\newcommand{\rsqrt}{\ensuremath{^{\frac{1}{2}}}}

\newcommand{\rHoFeDo}{HFD95}
\newcommand{\rWiNo}{WN86}
\newcommand{\rSiIcDo}{SID01}
\newcommand{\rBird}{B94}
\newcommand{\rKrGaSe}{KGS94}
\newcommand{\rKrSe}{KS97}

\begin{document}
\title{Three-component modeling of C-rich AGB star winds}
\subtitle{I. Method and first results}
\author{C.~Sandin\and{S.~H{\"o}fner}}
\institute{Department of Astronomy and Space Physics,
Uppsala University, Box 515, SE-751 20 Uppsala, Sweden}
\offprints{C.~Sandin,\\\email{Christer.Sandin@astro.uu.se}}
\date{Received /Accepted}

%%%%%%%%%%%%%%%%%%%%%%%%%%%%%%%%%%%%%%%%%%%%%%%%%%%%%%%%%%%%%%%%%%%%%%%%%%%%%%%
%%%%%%%%%%%%%%%%%%%%%%%%%%%%%%%%%%%%%%%%%%%%%%%%%%%%%%%%%%%%%%%%%%%%%%%%%%%%%%%
%%%%%%%%%      Abstract
%%%%%%%%%%%%%%%%%%%%%%%%%%%%%%%%%%%%%%%%%%%%%%%%%%%%%%%%%%%%%%%%%%%%%%%%%%%%%%%
%%%%%%%%%%%%%%%%%%%%%%%%%%%%%%%%%%%%%%%%%%%%%%%%%%%%%%%%%%%%%%%%%%%%%%%%%%%%%%%
\abstract{
Radiative acceleration of newly-formed dust grains and transfer of momentum from the dust to the gas plays an important role for driving winds of AGB stars. Therefore a detailed description of the interaction of gas and dust is a prerequisite for realistic models of such winds. In this paper we present the method and first results of a three-component time-dependent model of dust-driven AGB star winds. With the model we plan to study the role and effects of the gas-dust interaction on the mass loss and wind formation. The wind model includes separate conservation laws for each of the three components of gas, dust and the radiation field and is developed from an existing model which assumes position coupling between the gas and the dust. As a new feature we introduce a separate equation of motion for the dust component in order to fully separate the dust phase from the gas phase. The transfer of mass, energy and momentum between the phases is treated by interaction terms. We also carry out a detailed study of the physical form and influence of the momentum transfer term (the drag force) and three approximations to it. In the present study we are interested mainly in the effect of the new treatment of the dust velocity on dust-induced instabilities in the wind. As we want to study the consequences of the additional freedom of the dust velocity on the model we calculate winds both with and without the separate dust equation of motion. The wind models are calculated for several sets of stellar parameters. We find that there is a higher threshold in the carbon/oxygen abundance ratio at which winds form in the new model. The winds of the new models, which include drift, differ from the previously stationary winds, and the winds with the lowest mass loss rates no longer form.
}

\maketitle

%%%%%%%%%%%%%%%%%%%%%%%%%%%%%%%%%%%%%%%%%%%%%%%%%%%%%%%%%%%%%%%%%%%%%%%%%%%%%%%
%%%%%%%%%%%%%%%%%%%%%%%%%%%%%%%%%%%%%%%%%%%%%%%%%%%%%%%%%%%%%%%%%%%%%%%%%%%%%%%
%%%%%%%%%      Introduction
%%%%%%%%%%%%%%%%%%%%%%%%%%%%%%%%%%%%%%%%%%%%%%%%%%%%%%%%%%%%%%%%%%%%%%%%%%%%%%%
%%%%%%%%%%%%%%%%%%%%%%%%%%%%%%%%%%%%%%%%%%%%%%%%%%%%%%%%%%%%%%%%%%%%%%%%%%%%%%%
\section{Introduction}
% AGB atmosphere model background %%%%%%%%%%%%%%%%%%%%%%%%%%%%%%%%%%%%%%%%%%%%%
% Physical processes %%%%%%%%%%%%%%%%%%%%%%%%%%%%%%%%%%%%%%%%%%%%%%%%%%%%%%%%%
The extended atmospheres of AGB stars are sites of large local and global variations in physical quantities which reflect the variability of the stars. It is here that dust grains condense from the gas phase. The opaque dust is pushed out by the radiation pressure from the luminous star. The effects of interactions involving dust and shock waves caused by stellar pulsations can at the right conditions lead to the formation of a massive stellar wind. This process is critical for the evolution of AGB stars. The stellar wind does not only limit their lifetime but also enriches the surroundings with processed matter. First a circumstellar envelope is formed and later the products are mixed with the interstellar medium. Observed long-time variations in mass loss rates indicate that the properties of the stellar winds change as the stars evolve. The AGB phase ends when the star has almost completely lost its envelope and soon afterwards appears as a white dwarf surrounded by a planetary nebula.

Episodic mass loss variations of AGB stars on long time scales, i.e\@. $10^4$-$10^5$ years, have been observed in the form of detached CO shells for a long time~\citep[e.g.][]{OlBeErGu:96,OlBeLu.:00}. \citet{StSc:00} argue that thermal pulses likely are responsible for the origin of the detached CO shells. Their circumstellar envelope model includes separate equations of motion for gas and dust that are coupled to radiative transfer. In this context we also mention that there are variations on shorter time scales, on the order of $10^2$-$10^3$ years, that are believed to be associated with the duration of a helium shell flash at the beginning of a thermal pulse cycle.

Mass loss variations on time scales of $10^2$-$10^3$ years that unlikely can be explained by thermal pulses have more recently been observed in the form of concentric arcs~\citep[concentric shells; e.g.][and references therein]{MaHu:99,MaHu:00}. \citet[henceforth~{\rSiIcDo}]{SiIcDo:01} draw the conclusion that a two-fluid gas-dust interaction produces mass loss variations on a time scale of about $10^2$-$10^3$ years, that is seen to agree with observations of the dust-enshrouded star IRC +10216. Time dependent dust formation is included in their model.

From a different perspective another physical mechanism is proposed to play the key role in wind models of Late-AGB and Post-AGB objects. \citet{So:02} argues that the concentric arcs (M-arcs) observed around these objects are unlikely to originate in the wind acceleration zone through the interaction of gas and dust. Instead, they could be the result of an (ad hoc) solar-like magnetic activity cycle in the star~\citep{So:00}. \citet{GaLoFr:01} also find, without including the dynamic effects of the dust component, that a solar-like magnetic cycle without mass loss variations reproduces many properties of observed concentric arcs.

These studies motivate a closer investigation of the effects of the dust-gas coupling on AGB wind structures. Not only is a closer study of the origin of the shells interesting. From a more fundamental point of view the gas-dust coupling is \emph{essential} for the radiative driving of a stellar wind. It is important to study the limits of this coupling. Moreover, a model with improved physical capabilities will provide the grounds for both qualitatively and quantitatively better estimates of mass loss rates and spectral energy distributions. The conditions for stellar dust formation can also be better understood.

% Dynamical models: stationary & time-dependent %%%%%%%%%%%%%%%%%%%%%%%%%%%%%%%
To describe the wind correctly the models have to include a sufficient treatment of all three interacting components: gas, dust and the radiation field. Existing AGB wind models are either stationary or time-dependent. The models based on a stationary formulation do not admit flow variations with time. To their disadvantage few stellar parameter configurations have been shown to support stationary outflows (winds). On the other hand, time-dependent models tend to have (over-) simplified descriptions of radiative transfer (e.g\@. a semi-analytical treatment, or inadequate (often gray) opacities).

Another model subdivision can be made regarding the degree of coupling between the gas and the dust components. In models assuming complete momentum coupling all radiative momentum gained by the dust immediately is transferred to the gas. Position coupled models, in addition to complete momentum coupling, assume that the dust is mechanically bound to the gas phase, i.e\@. that it moves at the same velocity.

The latest group of models in the literature however do not put any of the mentioned restrictions on the dust velocity. The degree of coupling inevitably affects the physical distribution of both the gas and the dust in the envelope. However, without detailed modeling it is not clear quantitatively how large the effect due to the coupling will be. The most recent works concerning the influence of the treatment of the gas and dust phases have been carried out by~\citet{LiLaBe:01}, {\rSiIcDo} and~\citet{StSzSc:98}. An overview of the handling of the gas-dust interaction in earlier AGB wind models is presented by~{\rSiIcDo}.

% The work of this article %%%%%%%%%%%%%%%%%%%%%%%%%%%%%%%%%%%%%%%%%%%%%%%%%%%%
% Mapping %%%%%%%%%%%%%%%%%%%%%%%%%%%%%%%%%%%%%%%%%%%%%%%%%%%%%%%%%%%%%%%%%%%%%
In this work we describe the method of building a three-component AGB star wind model. This model uses, to our knowledge, the most complete time-dependent description of all three components, gas, dust and radiation, and it does not assume complete momentum coupling or position coupling. The physics and basic equations are presented in Sect.~\ref{sec:Phys}. The numerical method and a discussion of different numerical approximations of the phase interaction terms are given in Sect.~\ref{sec:Nume}. This work is a study of dust-induced dynamic instabilities in the atmosphere. We study the dynamics of three-component wind models and compare the results with corresponding models where position coupling, and hence complete momentum coupling, are assumed. The emphasis of the study is put on the effects of detailed momentum transfer. Section~\ref{sec:Resu} contains the results and a discussion while the conclusions are given in Sect.~\ref{sec:Conc}.

%%%%%%%%%%%%%%%%%%%%%%%%%%%%%%%%%%%%%%%%%%%%%%%%%%%%%%%%%%%%%%%%%%%%%%%%%%%%%%%
%%%%%%%%%%%%%%%%%%%%%%%%%%%%%%%%%%%%%%%%%%%%%%%%%%%%%%%%%%%%%%%%%%%%%%%%%%%%%%%
%%%%%%%%%      Physics
%%%%%%%%%%%%%%%%%%%%%%%%%%%%%%%%%%%%%%%%%%%%%%%%%%%%%%%%%%%%%%%%%%%%%%%%%%%%%%%
%%%%%%%%%%%%%%%%%%%%%%%%%%%%%%%%%%%%%%%%%%%%%%%%%%%%%%%%%%%%%%%%%%%%%%%%%%%%%%%
\section{Physics of the wind model}\label{sec:Phys}
\subsection{Original model characteristics}\label{sec:Phys_char}
The present work is based on an AGB wind model of~\citet[henceforth~{\rHoFeDo}]{HoFeDo:95}. We shall first summarize the properties of that model in this subsection and then discuss the modifications in Sect.~\ref{sec:Phys_duem}. The two-component (radiation hydrodynamic) RHD-system of the stellar model as described by~\citet{FeDoHo:93} was combined with dust as a third component by {\rHoFeDo}, defining the RHDD-system. Some assumptions made in the RHDD-system concerning the dust and the gas are important in the current work and require some explanation.

The matter in the wind is present in either of the two phases of dust or gas. The gas represents the overwhelmingly largest part of the matter (in some parts of the atmosphere there is even no dust at all). The hydrodynamic equations describing the gas phase are the equation of continuity, the equation of motion and the equation of (internal) energy. A perfect gas law is adopted for the equation of state; the ratio of the specific heats $\gamma=5/3$, and the mean molecular weight $\mu=1.26$.

Only those particles in the gas that are part of the dust chemistry can move between the two phases. The dust phase is assumed to be composed of spherical dust grains in the form of amorphous carbon. The dust equations that correspond to the equation of continuity for the gas phase (and describe the formation and destruction of dust grains) are the four moment equations for the moments $K_0$-$K_3$ of the grain size distribution function~\citep{GaSe:88,GaSeGa:90}. Dust formation is hereby treated self-consistently including the processes of nucleation, growth, evaporation and chemical sputtering (by gas particles) in a collision-less dust medium.
The moments are related to the average of powers of the dust grain radius and allow the calculation of average properties of the dust grains~\citep{GaKeSe:84} such as: the total number density of dust grains $n_\mathrm{d}=K_0$; the mean grain radius $\langle r_\mathrm{d}\rangle=r_0K_1/K_0$ (where $r_0$ is the monomer radius); the mean grain surface area $\langle A\rangle=4\pi r_0^2K_2/K_0$; the total number density of monomers condensed into grains $K_3$ gives the grain size $\langle N\rangle=K_3/K_0$; the dust mass density is $\rhod=m_1K_3$ ($m_1$ is the dust grain monomer mass). The number densities of the gas-phase molecules that are involved in the grain formation are calculated in an equilibrium chemistry of H, H$_2$, CO, C, C$_2$, C$_2$H and C$_2$H$_2$. The last four species contribute to the grain formation processes. All abundances are solar except for the carbon abundance which is specified through the carbon to oxygen ratio ($\varepsilon_{\rm C}/\varepsilon_{\rm O}$).

The model assumes complete momentum coupling; all momentum gained by the dust from the radiation field is immediately transferred to the gas. Assuming a very efficient mechanical (position) coupling of the dust to the gas~\citep{DoSeGa:89}, the two phases are defined to move at the same velocity. The resulting equation of motion is,
\begin{multline}
\frac{\partial}{\partial t}(\rho u)+\nabla\cdot(\rho u\,u)=\\
f_\mathrm{pressure,g}+f_\mathrm{grav,g}+f_\mathrm{rad,g}+f_\mathrm{rad,d}=\\
-\nabla P-\frac{Gm_r}{r^2}\rho+
  \frac{4\pi}{c}\kappa_\mathrm{g}\rho H+
  \frac{4\pi}{c}\kappa_\mathrm{d}\rho H\label{eq:Phys_PCeqmot}
\end{multline}
where $f_\mathrm{pressure,g}$ is the force due to the gradient of the gas pressure; $f_\mathrm{grav,g}$ is the gravitational force acting on the gas; $f_\mathrm{rad,g}$ is the radiative pressure force acting on the gas; $f_\mathrm{rad,d}$ is the radiative pressure force acting on the dust. Furthermore $\rho$ is the gas density; $u$ the gas (and dust) velocity; $\kappa_\mathrm{g}$ the (gray) gas opacity; $\kappa_\mathrm{d}$ the (gray) dust opacity; $P$ the gas pressure; $H$ the first moment of the radiation field. Note that since $\rho_\mathrm{d}\ll\rho$ -- and therefore $f_\mathrm{grav,d}\ll f_\mathrm{grav,g}$ -- it is assumed that the only important term related to the dust in this equation is the radiative pressure acting on the dust.

The dust (internal) energy equation is replaced by a radiative equilibrium relation since the dust is effectively thermally coupled to the radiation field. By assuming radiative equilibrium and LTE the dust temperature $T_\mathrm{d}$ is equal to the radiation temperature $T_\mathrm{rad}$ in the gray case. The radiation temperature is in turn defined by the radiative energy density $J$,
\begin{eqnarray}
J=\frac{\sigma_\mathrm{B}}{\pi}T_\mathrm{rad}^4
\end{eqnarray}
where $\sigma_\mathrm{B}$ is the Stefan-Boltzmann constant, $J$ is the zeroth moment of the radiation field. In addition the dust holds a negligible thermal energy compared to the radiative energy and the gas internal energy~\citep[cf.~section 2.2f in][and references therein]{HoDo:92}.

The radiation field is described by the frequency-integrated zeroth and first moments of the radiation intensity. These moments represent the radiative energy density and radiative energy flux respectively. The corresponding moment equations of the radiative transfer equation are solved together with the hydrodynamic equations for the gas and the dust. At each time-step the equation of radiative transfer is solved for a given structure using the method of characteristics~\citep[e.g.][]{Yo:80,Ba:88}. This solution makes it possible to calculate the Eddington factor and other quantities that are necessary to close the moment equations. The gas and dust opacities are both assumed to be gray. The gas opacity is set to~\citep{Bo:88},
\begin{eqnarray}
\kappa_\mathrm{g}=2\times10^{-4}\ [\mbox{cm}^2\ \mathrm{g}^{-1}]
\end{eqnarray}
while the dust opacity is~\citep[cf.][]{FlGaSe:92},
\begin{eqnarray}
\kappa_\mathrm{d}=\frac{\pi r_0^3}{\rho}Q'_\mathrm{ext}K_3,
\ \mbox{and}\ Q'_\mathrm{ext}=
\frac{Q_\mathrm{ext}}{\langle r_\mathrm{d}\rangle}
\end{eqnarray}
where $Q_\mathrm{ext}$ is the grain extinction efficiency. Previously, the Rosseland mean of $Q'_\mathrm{ext}$ was assumed as $Q'_\mathrm{ext}[\,\mbox{cm}^{-1}]=5.9\,T_\mathrm{d}$ (used in e.g.~{\rHoFeDo}; $T_\mathrm{d}$ is given in K). In later works (including this) it has been replaced with a better fit to the opacity data, $Q'_\mathrm{ext}[\,\mbox{cm}^{-1}]=4.4\,T_\mathrm{d}$~\citep[Winters 1994, priv.~comm.; cf\@.][]{HoDo:97}. The dependence of $Q'_{\mathrm{ext}}$ on $T_{\mathrm{d}}$ follows as a consequence of taking the Rosseland mean of the frequency-dependent extinction coefficient. The consistent treatment of the radiation field is a strength of the RHDD-system model compared to models using semi-analytical approximations. A more recent formulation of the RHDD-system also includes non-gray radiative transfer~\citep{Ho:99b,HoGaArJo:02,HoLoArJo:02}.\\

One physical limitation of the RHDD-system description in previous papers is the assumption of position coupling. By this assumption the effects of two drifting phases are totally disregarded, and can therefore not be properly considered. A separation of the two phases does not only require an additional equation of motion for the dust component but also several phase interaction terms.

The naming convention in the literature of AGB wind models varies depending on the description of the radiation field or the presence of a freely moving dust component. Models in which the focus is on dynamics are often called ``n-fluid'' models where the ``n'' denotes the number of separate equations of motion for different material components. We prefer to label the models according to for how many physical components the conservation laws are solved -- counting both material phases \emph{and} the radiation field -- emphasizing that we neither use simple equilibrium assumptions, nor prescribe values for a particular component which appear in interaction terms of other components. In this sense, we refer to our models as three-component models (gas, dust and radiation field), not as single- or two-fluid models. Furthermore, in our notation the RHDD-system models are \emph{position coupled} (PC) three-component models\footnote{In spite of the PC in these models the dust is still treated as a separate component, since the time-dependent formation, growth and evaporation of grains are described by the moment equations.} as opposed to the three-component \emph{drift models} that are described in the following subsection.

\subsection{The dust equation of motion}\label{sec:Phys_duem}
%%%%% Overview of earlier work
Most studies of dust-driven stellar winds have evolved around the assumption of complete momentum coupling in stationary winds. They also often contain a simplified description of either the radiation field or the dust component (e.g\@. instantaneous dust formation and a constant grain size). The two latest works on drift in stellar winds of AGB stars are those by~\citet{LiLaBe:01} and~{\rSiIcDo} (we refer to~{\rSiIcDo} for an overview of previous studies concerning the effects of drifting phases in cool stellar winds). The former have carried out a study of different degrees of dust-gas coupling in stationary winds of late-type stars using a frequency dependent dust opacity. However, the grain radius in the dust component is assumed constant. {\rSiIcDo} have carried out explicit time-dependent hydrodynamical modeling in a two-fluid medium using a given temperature structure, but treat dust formation in detail.

%%%%% Adding the new equation of motion
In this study we relax the assumptions of position coupling as well as complete momentum coupling made in the previous modeling (see Sect.~\ref{sec:Phys_char}). We do this by adding an equation of motion for the dust component to the RHDD-system and modifying the equation of motion for the gas accordingly. Phase interaction terms are included to conserve the physical quantities (see the following subsection). The resulting system is henceforth referred to as the RHD3-system, where the third ``D'' stands for drift; the models are accordingly named drift models.

The dust component consists of dust grains of different sizes. In principle each group of dust grains of a certain radius can be ascribed a separate equation of motion. Coupling terms between the equations for dust grains of different sizes may be neglected on the assumption that grain-grain collisions are far less frequent than grain-gas collisions (the dust is assumed to be pressure-less). We can then take a mean of the velocity equations of individual sizes to get the size-averaged dust equation of motion. This mean equation formally looks like an equation of motion for one grain size.

In the new system the gas equation of motion, Eq.~(\ref{eq:Phys_PCeqmot}), is exchanged with,
\begin{multline}
\frac{\partial}{\partial t}(\rho u)+\nabla\cdot(\rho u\,u)=\\
-\nabla P-\frac{Gm_r}{r^2}\rho+\frac{4\pi}{c}\kappa_\mathrm{g}\rho H
+f_{\mathrm{drag}}-\mathcal{S}_\mathrm{cond}u^\mathrm{i}\label{eq:Phys_eqmot}
\end{multline}
where the last two terms represent the momentum transfer by gas-dust collisions and the momentum change when dust condenses from the gas phase respectively (see Sect.\ref{sec:Phys_mass}). The pressure-free dust equation of motion in turn is,
\begin{multline}
\frac{\partial}{\partial t}(\rhod v)+\nabla\cdot(\rhod v\,v)=\\
-\frac{Gm_r}{r^2}\rhod+\frac{4\pi}{c}\kappa_\mathrm{d}\rho H-f_{\mathrm{drag}}+
\mathcal{S}_\mathrm{cond}u^\mathrm{i}\label{eq:Phys_eqdmot}
\end{multline}
$v$ is now the ``mean'' dust velocity and $\rhod$ is the dust density.

By the same reasoning as in the RHDD-system we do not include the dust equation of internal energy but assume that the grain temperature is determined by radiative equilibrium.

The dust formation processes are affected in several ways by drift~\citep[e.g.~\citealt{KrSe:97}, henceforth {\rKrSe};][]{DoSeGa:89,DrSa:79}. We do not include these modifications in the models presented in this paper, but plan to do it in the future.

\subsection{Phase interaction terms}\label{sec:Phys_pit}
In the gas-dust phase interaction we must consider each of the three transferred physical quantities of mass, momentum and internal energy. There are two types of momentum and energy exchange between the gas and dust phases. On the one hand they are transferred with the mass that switches phase, and on the other hand they are transferred in the collisional interaction.

\subsubsection{Mass transfer interaction terms}\label{sec:Phys_mass}
Mass is transferred between the phases when dust grains form or grow or alternatively when they evaporate. The rate at which material condenses $\mathcal{S}_\mathrm{cond}/m_1$ corresponds to the r.h.s\@. combination of source terms in the $K_3$ equation (Eq.~7 in {\rHoFeDo}). The rate $\mathcal{S}_\mathrm{cond}$ is also a sink term in the gas equation of continuity. The rate of the momentum transfer at mass transfer is represented by the last term in Eqs.~(\ref{eq:Phys_eqmot}~\&~\ref{eq:Phys_eqdmot}). We set the velocity of the formed (or evaporated) dust $u^\mathrm{i}$ equal to the gas velocity $u$.
The rate at which internal energy is transferred, and work is done by removal of mass from the gas phase, $\mathcal{S}_\mathrm{cond}h^\mathrm{n}$, is a sink term in the gas internal energy equation. Here $h^\mathrm{n}$ is the specific enthalpy,
\begin{eqnarray}
h^\mathrm{n}=e^\mathrm{n}+P^\mathrm{n}/\rho^\mathrm{n}\,.
\end{eqnarray}
The superscript n indicates that, on addition of mass to the gas, these quantities might be in non-equilibrium with the gas.

The effects of the  mass transfer on the gas are very small compared to the effects on the dust. Nevertheless we include all terms above in both the dust and the gas equations for completeness, with the exception of the internal energy transfer term ($\mathcal{S}_\mathrm{cond}h^\mathrm{n}$). The internal energy transfer to (and from) the gas phase is ignored on the basis that the internal energy transfer with the radiative field always will be (several) orders of magnitude larger.

In contrast to the terms discussed next all interaction terms mentioned so far are independent of the presence of the separate dust equation of motion

\subsubsection{Collisional interaction terms}\label{sec:Phys_inte}
The momentum transfer term, the drag force, is the most important interaction term in dust-driven winds. Most of the radiative pressure acts on the opaque dust grains accelerating them outwards. The gas is dragged along by the accelerating dust and forms a stellar wind.

The drag force is derived from the local physical conditions. In our case these can be characterized as follows. The gas particle velocities have a Maxwellian distribution (the gas is described with the continuum approximation). The dust grains in the dust medium are in the free-molecular regime compared to the gas. In App.~\ref{sec:Appe_cond} we discuss the validity of this and the previous assumption in the current context. The dust grains are assumed to be spherical.  
Collisions between gas particles and the dust grain surface can be either specular or diffusive, depending on how the normal and tangential momentum is distributed in the collision. In a specular collision of the incident particle the normal component of the velocity (in the frame of reference of the dust particle) is reversed on reflection, while the tangential components are unaffected. In a diffusive collision the particle is first acommodated on the surface, then it is thermalized, and finally it is emitted in a random direction. The drag force is derived by integrating the pressure over the surface of the dust grain~\citep[cf\@. e.g.][]{HaPr:59,Sc:63}. The general form of the drag force is
\begin{eqnarray}\label{eq:Phys_fdrag}
\fdrag=\sigma\rho n_\mathrm{d}\frac{v_\mathrm{D}^2C\ssD}{2}
\end{eqnarray}
where $\sigma$ denotes the gas-dust geometrical cross section; $v_\mathrm{D}$ is the drift velocity ($v-u$); $C_\mathrm{D}$ is the drag coefficient. The factor of 2 comes from the definition of the drag coefficient. For $\sigma$ we use,
\begin{eqnarray}
\sigma=\pi\langle r_\mathrm{d}\rangle^2=\pi r_0^2K_1^2/K_0^2\,.
\end{eqnarray}
Because of its conciseness we prefer the form of the drag coefficient presented by~\citet[henceforth {\rBird}]{Bi:94},
\begin{eqnarray}
C\ssD=C_{\mathrm{D},\mathrm{diff}}+C_{\mathrm{D},\mathrm{common}}=
\frac{2}{3}\frac{\pi\rsqrt(1-\varepsilon)}{S_\mathrm{D}}
\left(\frac{T_\mathrm{d}}{T_\mathrm{g}}\right)\rsqrt+\nonumber\\
+\left[\frac{4S_\mathrm{D}^4+4S_\mathrm{D}^2-1}{2S_\mathrm{D}^4}\mathrm{erf}(S_\mathrm{D})
+\frac{2S_\mathrm{D}^2+1}{\pi\rsqrt S_\mathrm{D}^3}\exp(-S_\mathrm{D}^2)\right]\label{eq:Phys_CDanalytic}
\end{eqnarray}
where erf is the error function; $T_\mathrm{g}$ is the gas kinetic temperature; the fraction of specular collisions is defined by $\varepsilon$ (see below); $S_\mathrm{D}$ is the speed ratio,
\begin{eqnarray}
S_\mathrm{D}=\frac{v\ssD}{v_\mathrm{mp}},\ \mbox{where}\ v_\mathrm{mp}=
\sqrt{\frac{2k_\mathrm{B}T_\mathrm{g}}{\mu m_\mathrm{H}}}\,.
%\label{eq:Phys_S}\,.
\end{eqnarray}
$v_\mathrm{mp}$ is the most probable thermal speed of the Maxwellian velocity distribution.

The second term on the r.h.s\@. in Eq.~(\ref{eq:Phys_CDanalytic}), $C_{\mathrm{D},\mathrm{common}}$ (the common term), is sufficient when all collisions are assumed to be specular. The first term, $C_{\mathrm{D},\mathrm{diff}}$, is in addition required in diffusive collisions. $\varepsilon=1$ corresponds to fully specular collisions and $\varepsilon=0$ to fully diffusive collisions. A combination of collisions is achieved by using values on $\varepsilon$ in between (however as pointed out by {\rBird} this combination is only an approximation to a real scattering law). It is not yet established how collisions are distributed between specular and diffusive~\citep[see however e.g.][henceforth {\rKrGaSe}]{KrGaSe:94}, and we leave the option to use a combination of them.

In the collisional interaction the internal energy of the gas is modified by on the one hand inelastic (diffusive) dust-gas collisions (cf\@. the discussion on `$q_\mathrm{acc}$' in {\rKrGaSe}). On the other hand kinetic energy is converted into internal (thermal) energy as gas particles which preferentially come from one direction (the drift) are reflected in random directions when hitting a dust particle. The energy transferred to the gas in this process, in addition to the work done by the drag force on the gas, corresponds (in the case of only specular collisions) to the relative speed of the gas and dust particles times the drag force (cf\@. the treatment of `$q_\mathrm{fric}$' in {\rKrGaSe}). We ignore the effects of both these terms ($q_\mathrm{acc}$ and $q_\mathrm{fric}$) in the models presented in this article, on the assumption that the effects on the gas internal energy are minute when compared to the radiative energy exchange with the gas. Preliminary results of our time-dependent models including the second heating term ($q_\mathrm{fric}$) show that its influence on the structure is negligible.

In connection to this discussion we want to point out that the same term was found by {\rKrGaSe} to play a significant role in the gas energy balance in their stationary models. It is, however, difficult to compare their results with ours because the models differ in several respects. The most important differences are: different physical assumptions on the radiative transfer; different stellar parameters of the models; they use constant sized dust grains (i.e\@. no time-dependent formation). We plan to discuss the importance of this term in a forthcoming article.

\subsection{Complete momentum coupling}
The validity of the assumption of complete momentum coupling (CMC) in the cool stellar dust-driven wind has been subject of extensive discussions starting with~\citet{Gi:72}. For CMC flows it is assumed that all radiation momentum is immediately transferred to the gas, and the inertia of the dust phase is neglected. Hence,
\begin{eqnarray}
\fdrag=f_\mathrm{rad,d}+f_\mathrm{grav,d}\label{eq:Phys_CMC}
\end{eqnarray}
where $f_\mathrm{grav,d}$ is the gravitational force acting on the dust. With only specular collisions -- and using one of the approximative forms of the drag coefficient $C_\mathrm{D}$ for $\fdrag$ presented in Sect.~\ref{sec:Nume_RHD3} -- this expression can be inverted with respect to the (equilibrium) drift velocity $\overline{v\ssD}$~(see e.g\@. section 2.2.3 {\rKrSe}). With this relation there is no need to solve the dust equation of motion, and computational time is saved when solving the system of equations without it. However when we consider diffusive collisions an inversion is not possible anymore and Eq.~(\ref{eq:Phys_CMC}) must be solved numerically. In that case there is no computational motivation to use the assumption of CMC, and one may as well solve the dust equation of motion. We do not assume CMC in the RHD3-system but we give a qualitiative criterion on how close our models are to CMC in Sect.~\ref{sec:Resu_mome}. Equilibrium drift expressions are used by e.g\@. {\rKrSe} and in some of the calculations presented in {\rSiIcDo}.

%%%%%%%%%%%%%%%%%%%%%%%%%%%%%%%%%%%%%%%%%%%%%%%%%%%%%%%%%%%%%%%%%%%%%%%%%%%%%%%
%%%%%%%%%%%%%%%%%%%%%%%%%%%%%%%%%%%%%%%%%%%%%%%%%%%%%%%%%%%%%%%%%%%%%%%%%%%%%%%
%%%%%%%%%      Numerical method
%%%%%%%%%%%%%%%%%%%%%%%%%%%%%%%%%%%%%%%%%%%%%%%%%%%%%%%%%%%%%%%%%%%%%%%%%%%%%%%
%%%%%%%%%%%%%%%%%%%%%%%%%%%%%%%%%%%%%%%%%%%%%%%%%%%%%%%%%%%%%%%%%%%%%%%%%%%%%%%
\section{Numerical method}\label{sec:Nume}
%%%%% RHDD-system %%%%%%%%%%%%%%%%%%%%%%%%%%%%%%%%%%%%%%%%%%%%%%%%%%%%%%%%%%%%%
Before we look at the numerical differences of the RHD3-system compared to the RHDD-system we summarize the main features of the RHDD-system.

\subsection{Features of the RHDD-system}\label{sec:Nume_RHDD}
A detailed description of the numerical method of the RHDD-system is found in~\citet{DoHo:91}. The RHD-system, without dust, was described in~\citet{DoFe:95,DoFe:91}.

% -- Grid equation
The gridpoints are distributed with an adaptive grid~\citep{DoDr:87} in which a grid equation resolves gradients of selected quantities. Currently the grid resolution function is determined by the thermal (internal) energy and the gas density. The temporal smoothing factor is set to $\tau_\mathrm{g}=10^2\ \mbox{s}$, which is orders of magnitude smaller than, e.g., dynamical or dust time scales in the problem, meaning that the grid can freely adapt to physical features; the spatial smoothing factor is set to $\alpha=2$.

% -- Artificial viscosity
Artificial tensor viscosity~\citep{TsWi:79} is used in regions subject to inhomologous contraction. The term is added as a source term in the gas equation of motion and the (internal) energy equation. The shock front is thereby widened to the relative characteristic length scale $l$,
\begin{eqnarray}
l=rf\label{eq:Nume_avg}
\end{eqnarray}
where $r$ is the local scale length, i.e\@. the radial distance from the center, and $f$ is a constant that defines the width of the shock front as a fraction of $r$.

In addition to the five RHD-equations, the four dust moment equations and the grid equation there are two more equations; an equation of the integrated mass and an equation keeping track of the condensible amount of carbon. Thus totally there are twelve non-linear equations, out of which ten are partial differential equations (PDEs). All equations are discretized in the volume-integrated conservation form on a staggered mesh~\citep[henceforth~\rWiNo]{WiNo:86}. The spatial discretization of the advection term can be chosen to be either first order (donor cell), or second order~\citep[monotonic advection, ][]{vLe:77}. The same order of precision is used in all PDEs.

% -- Implicit solution of the system using a modified Newton-Raphson technique
The full RHDD-system of twelve equations is solved implicitly using a Newton Raphson algorithm where the Jacobian of the system is inverted by the Henyey method.

%%%%% RHD3-system %%%%%%%%%%%%%%%%%%%%%%%%%%%%%%%%%%%%%%%%%%%%%%%%%%%%%%%%%%%%%
\subsection{Numerical issues in the RHD3-system}\label{sec:Nume_RHD3}
In this section we address several numerical issues associated with the dust equation of motion. The RHD3-system now consists of thirteen equations.

The grid equation can be adjusted to resolve both the gas and the dust components by including corresponding dust quantities in the grid resolution function~\citep[as suggested by][]{DoGa:90}. If this is done, however, the number of gridpoints should be increased to resolve both components. An increased number of gridpoints has the disadvantage that the fraction of mass contained in some grid cells may become smaller than the numerical accuracy of the scheme, thereby introducing new problems. Hence we leave the grid resolution function unchanged (compared to {\rHoFeDo}). Presently we use 500 gridpoints in all calculations, and a first order donor cell advection is adopted in all drift models.

An artificial viscosity term analogous to the one in the gas equation of motion is added to the dust equation of motion. Like the gas shocks, strong dust velocity gradients are widened by a characteristic dust front length scale (see Sect.~\ref{sec:Resu_thre} for a description of the term \emph{dust front}), i.e.
\begin{eqnarray}
l_\mathrm{d}=rf_\mathrm{d}\label{eq:Nume_avd}
\end{eqnarray}
We set both $f$ and $f_\mathrm{d}$ equal to $3.5\, 10^{-3}$ in all our calculations.

In drift models dust density gradients at shock fronts may become extremely steep, despite the widening achieved with the artificial viscosity. To smear out these gradients we add artificial diffusion in these models in the form presented by~{\rWiNo}. This term is added as a source term in all four dust moment equations,
\begin{eqnarray}
D_{K_i}\equiv\nabla\cdot\left(\varsigma_{K}
   \nabla K_i\right)\label{eq:Nume_amdi}
\end{eqnarray}
where $K_i$ denote the dust moments $K_0$-$K_3$. The transport coefficient $\varsigma$ is in general defined as,
\begin{eqnarray}
\varsigma=l^2/\tau
\end{eqnarray}
where $l$ is defined as above. We define the characteristic shock propagation time $\tau$, which is the time needed to cross a region of relative width $\Delta x$ as, $\tau=\Delta x/|w|$, where $w$ is the shock velocity. With the dust velocity $v$ as a measure of the shock velocity and the local scale length $r$ representing the relative width $\Delta x$ we have,
\begin{eqnarray}
\varsigma_{K}=f_\mathrm{d}^2r|v|\,.
\end{eqnarray}

A physically motivated explanation for the use of artificial mass diffusion in our drift models can be found in the experience from other areas of numerical hydrodynamics. As {\rWiNo} (and references therein) discuss, spurious results occur when strong shocks interact with walls, contact discontinuities and other strong shocks. Normally, there is enough numerical diffusion implicit in the advection of the numerical scheme to prevent these features from appearing, but if and when there is not they may appear as strange spikes in the solution.

In our drift models features (spikes) that can be attributed to the interaction between steep gas shocks and dust fronts appear first in the dust velocity, and subsequently in the dust density (i.e\@. the dust moments). Furthermore the time steps are decreased by the large variations of these quantities. Artificial diffusion prevents most of these features from appearing, but does not manage to remove all of them (see Sect.~\ref{sec:Resu_thre}). The model evolution and the wind characteristics are, however, not affected by their presence since the dust density in the region of the feature always is very small. While the described features are interpreted by the artificial viscosity term as contracting regions on the outwards facing side (i.e\@. away from the star), they are not interpreted as such on the inside. Consequently, these features can always be identified by the ``discontinuous'' jump on the inwards facing side. When we make estimates of the maximum dust speed, dust density variations and related quantities we must first separate these features from the rest of the structure.\\

% -- Dust equation of motion
The dust equation of motion (Eq.~\ref{eq:Phys_eqdmot}) is not numerically well defined in regions with very small amounts of dust. On the other hand, very small amounts of dust are not likely to affect the structure of the stellar wind. The dust moments become irrelevant for low degrees of condensation ($f_{\mathrm{cond}}$) that satisfy,
\begin{eqnarray}
f_\mathrm{cond}=
  \frac{\rhod}{\rho_\mathrm{c}^{\mathrm{tot}}}\approx
\frac{K_3}{K_3+n_\mathrm{c}}\lesssim\epsilon=f^\mathrm{min}_\mathrm{cond}
\end{eqnarray}
where $\rho_\mathrm{c}^{\mathrm{tot}}$ is the total density of condensible matter (both that present in the gas and the dust phases); $n_\mathrm{c}$ is the total number density of condensible material in the gas phase; $\epsilon$ is the numerical limit of an insignificant amount of dust. From our numerical experience we have found $\epsilon$ to be about $10^{-7}$. The total abundance of atoms of condensible material is about $10^{-4}$ of the total number of gas particles (this is the carbon that is not bound to CO). Thus for carbon,
\begin{eqnarray}
\frac{\rho_\mathrm{c}^{\mathrm{tot}}}{\rho}\approx12\times10^{-4}\sim10^{-3}\,.
\end{eqnarray}
We find the numerically limiting quantity of ``no dust'' to be,
\begin{eqnarray}
\left.\frac{\rhod}{\rho}\right|_{\mathrm{no}\ \mathrm{dust}}=
  \frac{\rho_\mathrm{c}^{\mathrm{tot}}}{\rho}
\cdot f_\mathrm{cond}^\mathrm{min}\approx10^{-3}\cdot10^{-7}=
10^{-10}\label{eq:Nume_duli}\,.
\end{eqnarray}
In regions where there is dust and the dust/gas ratio eventually becomes lower than the limit given above we exchange the dust equation of motion, Eq.~(\ref{eq:Phys_eqdmot}), with the relation,
\begin{eqnarray}
v=u\,.
\end{eqnarray}
For reasons of stability we use this relation also in dust forming regions where the dust/gas ratio is smaller than 10 times the value in Eq.~(\ref{eq:Nume_duli}).

\begin{figure}
  \resizebox{\hsize}{!}{\includegraphics{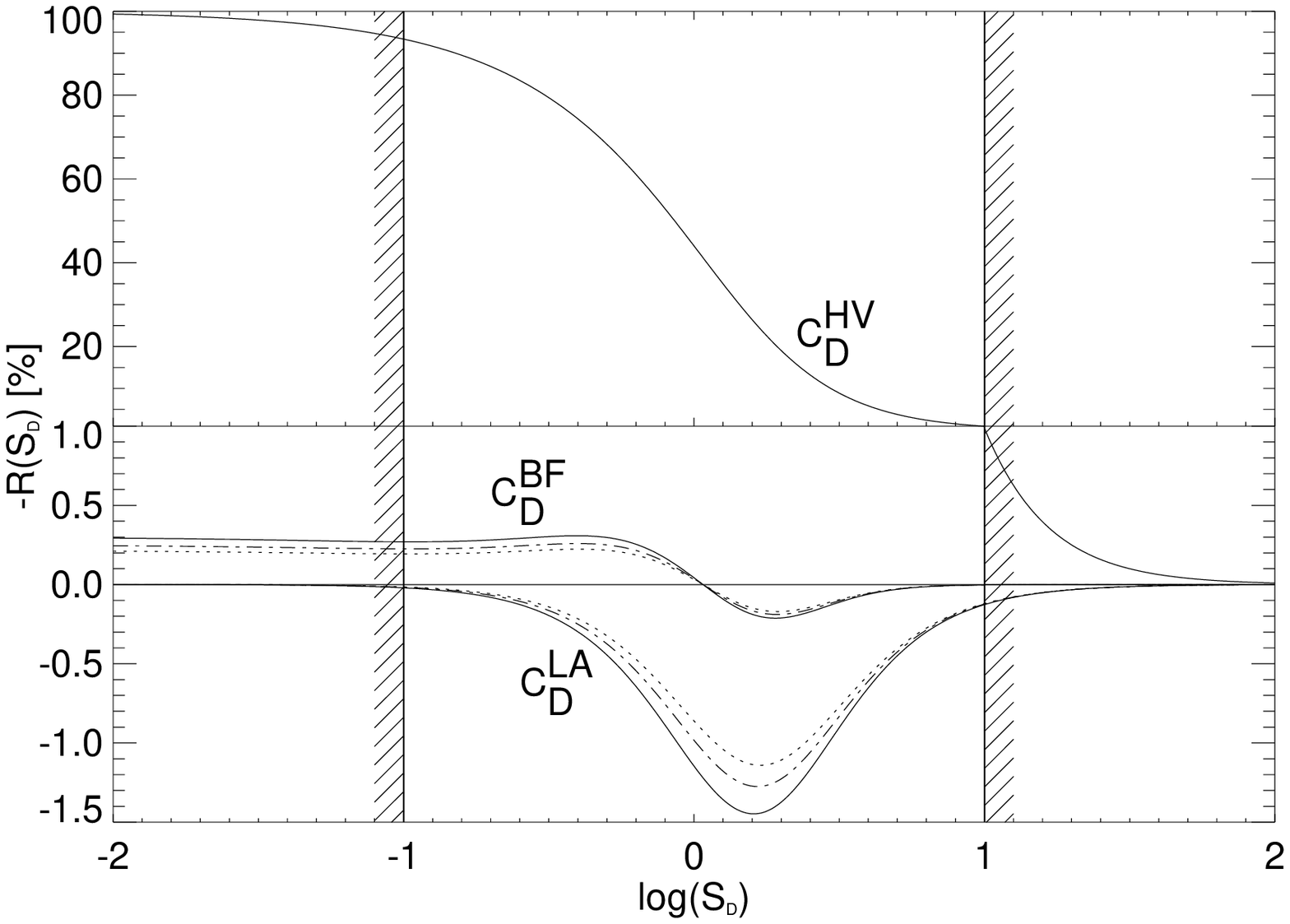}}
  \caption{Relative error of the approximative drag coefficients with respect to the expression $C\ssD$ (Eq.~\ref{eq:Phys_CDanalytic}). The limits approximation (Eq.~\ref{eq:Nume_CDSQ}) is represented by the three lines grouped at $C^{\mathrm{LA}}\ssD$; the approximation of~\citet{BeFr:83} (Eq.~\ref{eq:Nume_CDBF}) is represented by the three lines at $C^{\mathrm{BF}}\ssD$; the single line $C^{\mathrm{HV}}\ssD$ represents the high-velocity approximation (Eq.~\ref{eq:Nume_CDHV}). The discontinuity in $C^{\mathrm{HV}}\ssD$ at (1,1) is due to the discontinuous plot ordinate. Solid lines are computed assuming fully specular collisions ($\varepsilon=1$). The dotted lines assume fully diffusive collisions ($\varepsilon=0$). For the dash-dotted lines 50\% of the collisions ($\varepsilon=0.5$) are specular and the rest diffusive. The temperature ratio adopted in the diffusive term $C_{\mathrm{D},\mathrm{diff}}$ is $1.0$. The vertical bars show the range of typical values of the speed ratio $S_\mathrm{D}$ in our wind models (see Eq.~\ref{eq:Resu_S}).}
  \label{fig:Nume_CD}
\end{figure}

% -- Momentum transfer term
The part of the analytical drag coefficient common to both specular and diffusive collisions, $C_{\mathrm{D},\mathrm{common}}$ (Eq.~\ref{eq:Phys_CDanalytic}), is numerically difficult for small speed ratios $S_\mathrm{D}$. \citet{BaWiAs:65}, \citet{BeFr:83} and others have derived approximations to this expression. Here we compare three approximations of the drag coefficient with the expression in Eq.~\ref{eq:Phys_CDanalytic}. All three approximations are written below in the same form as $C\ssD$. We have added the diffusive collisions term to the first two approximations. \citeauthor{BeFr:83} derived the following approximation of $C_{\mathrm{D},\mathrm{common}}$~\citep[see also,][]{LiLaBe:01,MaMoCa:96},
\begin{eqnarray}
C^{\mathrm{BF}}\ssD=\frac{2}{3}\frac{\pi\rsqrt(1-\varepsilon)}{S_\mathrm{D}}
\left(\frac{T_\mathrm{d}}{T_\mathrm{g}}\right)\rsqrt+
        \frac{2}{S_\mathrm{D}^2}\mathrm{sgn}(S_\mathrm{D})\cdot\nonumber\\
\cdot\left[1+S_\mathrm{D}^2-
\frac{1}{(0.17S_\mathrm{D}^2+0.5|S_\mathrm{D}|+1)^3}\right]\,.
\label{eq:Nume_CDBF}
\end{eqnarray}
In the \emph{limits approximation} the separate solutions of high and low $S_\mathrm{D}$ are combined~\citep[see e.g\@. {\rKrGaSe},][]{McGSt:92,DrSa:79},
\begin{eqnarray}
C^{\mathrm{LA}}\ssD=\frac{2}{3}\frac{\pi\rsqrt(1-\varepsilon)}{S_\mathrm{D}}
\left(\frac{T_\mathrm{d}}{T_\mathrm{g}}\right)\rsqrt
+\frac{2}{S_\mathrm{D}}\left(\frac{64}{9\pi}+S_\mathrm{D}^2\right)\rsqrt
\label{eq:Nume_CDSQ}\,.
\end{eqnarray}
The influence of the thermal conditions is ignored in the \emph{high-velocity approximation}~(see e.g\@. {\rSiIcDo}),
\begin{eqnarray}
C^{\mathrm{HV}}\ssD=2\label{eq:Nume_CDHV}\,.
\end{eqnarray}
Figure~\ref{fig:Nume_CD} shows the relative errors of the approximative drag coefficients presented above compared to $C\ssD$. In the figure we see that $C^{\mathrm{BF}}\ssD$ deviates the least from $C\ssD$. The relative error of $C^{\mathrm{LA}}\ssD$ is never larger than 1.5\% and it is easier to code than $C^{\mathrm{BF}}\ssD$ and therefore we use $C^{\mathrm{LA}}\ssD$ in our calculations.

In their work on a two-component wind model~{\rSiIcDo} used the high-velocity approximation of the drag coefficient in the momentum transfer. This choice was motivated by the fact that the momentum transfer term makes the system of equations too stiff to be solved explicitly, thereby requiring very small time steps. To make an analytical time-integration of the drag force possible (and thereby allow larger time steps) they chose the simpler form achieved with $C^{\mathrm{HV}}\ssD$. In Sect.~\ref{sec:Resu_mome} we study the response of the RHD3-system to the two different drag coefficients $C^{\mathrm{HV}}\ssD$ and $C^{\mathrm{LA}}\ssD$.

The advantage of an implicit scheme is that the time step is not limited by the very restrictive Courant-Friedrichs-Lewy (CFL) condition. In our models it is instead limited by the temporal variations in the physical quantities. Currently local variations less than 20\% between two subsequent time steps are accepted or else the time step is reduced. Our experience shows that the dust velocity has the strongest variations of the quantities and therefore limits the time steps in drift models. The variations of the dust velocity are the largest in gridpoints containing very little dust mass. A negative effect of a shorter time step is that variations on shorter time scales (noise) may occur in other quantities as well. Slightly more than ten times as many time-steps are needed to reach the same age in drift models compared to PC models.

Note that the results presented here for the position coupled RHD3-system models do not exactly match the results of~{\rHoFeDo}. The deviations are caused by a modified dust opacity (see Sect.~\ref{sec:Phys_char}) and a few minor changes in how the dust moment equations are activated when the first dust grains form close to the photosphere.

%%%%%%%%%%%%%%%%%%%%%%%%%%%%%%%%%%%%%%%%%%%%%%%%%%%%%%%%%%%%%%%%%%%%%%%%%%%%%%%
%%%%%%%%%%%%%%%%%%%%%%%%%%%%%%%%%%%%%%%%%%%%%%%%%%%%%%%%%%%%%%%%%%%%%%%%%%%%%%%
%%%%%%%%%%%%%%%%%%%%%%%%%%%%%%%%%%%%%%%%%%%%%%%%%%%%%%%%%%%%%%%%%%%%%%%%%%%%%%%
%%%%%%%%%      Results, discussion
%%%%%%%%%%%%%%%%%%%%%%%%%%%%%%%%%%%%%%%%%%%%%%%%%%%%%%%%%%%%%%%%%%%%%%%%%%%%%%%
%%%%%%%%%%%%%%%%%%%%%%%%%%%%%%%%%%%%%%%%%%%%%%%%%%%%%%%%%%%%%%%%%%%%%%%%%%%%%%%
%%%%%%%%%%%%%%%%%%%%%%%%%%%%%%%%%%%%%%%%%%%%%%%%%%%%%%%%%%%%%%%%%%%%%%%%%%%%%%%
\section{Results and discussion}\label{sec:Resu}

%%%%%%%%%%%%%%%%%%%%%%%%%%%%%%%%%%%%%%%%%%%%%%%%%%%%%%%%%%%%%%%%%%%%%%%%%%%%%%%
%%%%%%%%%%%%%%%%%%%%%%%%%%%%%%%%%%%%%%%%%%%%%%%%%%%%%%%%%%%%%%%%%%%%%%%%%%%%%%%
%%%%%%%%%% Modeling procedure %%%%%%%%%%%%%%%%%%%%%%%%%%%%%%%%%%%%%%%%%%%%%%%%%
%%%%%%%%%%%%%%%%%%%%%%%%%%%%%%%%%%%%%%%%%%%%%%%%%%%%%%%%%%%%%%%%%%%%%%%%%%%%%%%
%%%%%%%%%%%%%%%%%%%%%%%%%%%%%%%%%%%%%%%%%%%%%%%%%%%%%%%%%%%%%%%%%%%%%%%%%%%%%%%
\subsection{Modeling procedure}\label{sec:Resu_mode}
The basic modeling procedure of all winds in this article is similar to that in {\rHoFeDo}. Note however that we use the RHD3-system code in all our calculations, including the PC models.

The modeling procedure is as follows. All wind models are started from hydrostatic dust-free initial models where the outer boundary is located at about $2\,R_*$. All five dust equations are switched on at the same time and dust starts to form whereby an outward motion of the dust and the gas is initiated. The expansion is followed by the grid to about $25\,R_*$ (typically $1\,10^{15} \mbox{cm}$). At this radius the outer boundary is fixed allowing outflow, and the model evolves for about 100 years (and more). The model calculation is stopped before a significant depletion of the mass inside the computational domain occurs.

In the models presented here it is the dust that initiates and drives the stellar wind. The dust formation zone is always inside the model domain. An inflow of matter through the inner boundary which is located well below the photosphere (typically at about $0.9\,R_*$) is not allowed. The fraction of the mass of the star contained in the model domain is about 15-30\%, depending on the stellar parameters. This large fraction is a consequence of using a rather low gas opacity for the present models (Sect.~\ref{sec:Phys_char}), and the problem was discussed by~\citet[cf\@. Sect.~3.1 and Fig.~1]{HoJoLoAr:98}. Note that a too high density in the stellar photosphere does not necessarily translate into unrealistic conditions in the wind acceleration zone since the density structure in this region is strongly influenced by dynamical effects. A smaller mass fraction is achieved by using a more appropriate description of the gas opacity, which we plan to do in a forthcoming article. We do not make specific assumptions on the presence of drift between the gas and dust phases in different parts of the atmosphere.

The model is determined by the three stellar parameters of stellar mass $M$, luminosity $L$ and the effective temperature $T_{\rm eff}$. In addition the carbon abundance is specified through the carbon/oxygen ratio $\varepsilon_{\rm C}/\varepsilon_{\rm O}$. The model parameters in our study are given in Table~\ref{tab:Resu_models}. The parameters of the models we study in this section match those of {\rHoFeDo}, for clarity the models are named with the corresponding letters.

The discretization of the advection terms of the PC models are either first or second order in precision while those in the the drift models, always are of first order precision (see Sect.~\ref{sec:Nume_RHD3}). To estimate the possible influence of 1st vs\@. 2nd order numerics on the model we present results using both alternatives in the PC models.

\begin{table}
\caption{Stellar parameters of the models. The models in this article are named in correspondence with the models in {\rHoFeDo}, Table~2.}
\begin{tabular}{ccrcrc}\hline\\[-1.8ex]
model & $M_*\ [M_{\sun}]$ & $L_*\ [L_{\sun}]$ &
        $T_\mathrm{eff}\ [\mbox{K}]$ & $R_*\ [R_{\sun}]$
\\[1.0ex]\hline\\[-1.8ex]
A & 1 & $10^4$     & 2600 & 493 \\
B & 1 & $10^4$     & 2500 & 533 \\
C & 1 & $10^4$     & 2400 & 578 \\
D & 1 & $3\, 10^4$ & 2600 & 853
\\[1.0ex]\hline%\\[-1.8ex]
\end{tabular}
\label{tab:Resu_models}
\end{table}

%%%%%%%%%%%%%%%%%%%%%%%%%%%%%%%%%%%%%%%%%%%%%%%%%%%%%%%%%%%%%%%%%%%%%%%%%%%%%%%
%%%%%%%%%%%%%%%%%%%%%%%%%%%%%%%%%%%%%%%%%%%%%%%%%%%%%%%%%%%%%%%%%%%%%%%%%%%%%%%
%%%%%%%%%%  Decoupled models vs. position-coupled, larger comparison %%%%%%%%%%
%%%%%%%%%%%%%%%%%%%%%%%%%%%%%%%%%%%%%%%%%%%%%%%%%%%%%%%%%%%%%%%%%%%%%%%%%%%%%%%
%%%%%%%%%%%%%%%%%%%%%%%%%%%%%%%%%%%%%%%%%%%%%%%%%%%%%%%%%%%%%%%%%%%%%%%%%%%%%%%
\subsection{Three-component dust-driven wind models}\label{sec:Resu_thre}
In this section we study the physical differences of drift models compared to PC models. In selecting the parameters for the models we wanted to cover both less massive and massive winds. What we describe is a dust-driven instability. Presently there are no pulsations since the use of a variable inner boundary condition that simulates pulsations could complicate the interpretation of other effects. Without an atmospheric levitation by (pulsational) shocks the models require higher $\varepsilon_{\rm C}/\varepsilon_{\rm O}$ ratios to form winds. The results of the models are given in Table~\ref{tab:Resu_thre}. Before we look at the results, we want to review the characteristics of the different winds that form.

Stationary winds by definition do not show variations. In PC models it has been found that low mass loss winds, with low  $\varepsilon_{\rm C}/\varepsilon_{\rm O}$ ratios, can become stationary. With higher $\varepsilon_C/\varepsilon_O$ ratios the dust formation process becomes unstable and the models consequently time-dependent. In the back-warmed region behind a newly formed dust shell of such a model the temperature eventually becomes too high for further dust formation to take place, and the process is shut off. Radiation pressure meanwhile acts on the individual grains to push the dust shell (and the gas) outwards. When the gas temperature behind the leaving dust shell has decreased sufficiently the dust formation process is reactivated and a new dust shell can form. This mechanism was called dust induced $\kappa$-mechanism by~\citealt{FlGaSe:95} and {\rHoFeDo}.

The conditions of wind formation change in drift models. The dust tends to accumulate in regions where the gas-dust interaction is strong. Strong interaction regions are typically represented by the dense gas behind shocks. When the interaction is not strong enough for the dust to drag the gas along, dust alone leaves the envelope and no significant wind forms. The dust component is a pressure-less medium that does not form ordinary shocks. However, dust tends to accumulate in high (gas) density regions which usually occur behind gas shocks. The result of the accumulation are strong gradients in the dust density and the dust velocity that coincide with the gas shocks. We will hereafter refer to these gradients as ``dust fronts''.

\begin{table*}
\caption{Model quantities for the PC models (2nd- and 1st-order precision in the advection terms; suffixes c2 and c1) and the drift models (suffix d1). The model names are constructed by combining the symbol in Table~\ref{tab:Resu_models} representing the combination of stellar parameters with ($10\times$) the carbon/oxygen ratio. The models that we were not able to run for a longer time interval are marked with parentheses; the associated numbers are less reliable because the means are taken over small time intervals. The different types of winds are: {\bf s}, stationary wind; {\bf i}, irregular wind; {\bf q}, periodic or quasi-periodic wind; {\bf ---}, no wind.}
\begin{tabular}{c|ccc|ccc|ccc}\hline
& \multicolumn{6}{c|}{} & & &\\
      & \multicolumn{6}{c|}{{\bf position coupled models}} & \multicolumn{3}{c}{{\bf drift models}}\\
& & & & & & & & &\\
      & \multicolumn{3}{c|}{2nd order (c2)} & \multicolumn{3}{c|}{1st order (c1)} & \multicolumn{3}{c}{1st order (d1)} \\
model,$\varepsilon_\mathrm{C}/\varepsilon_\mathrm{O}$
        & Type & $\langle\dot{M}\rangle$ & $\langle u_{\infty}\rangle$ &
          Type  & $\langle\dot{M}\rangle$ & $\langle u_{\infty}\rangle$ &
          Type  & $\langle\dot{M}\rangle$ & $\langle u_{\infty}\rangle$ \\
      & &                       $[\mbox{M}_{\sun}\mbox{yr}^{-1}]$ & $[\mbox{km}\,\mbox{s}^{-1}]$
      & & $[\mbox{M}_{\sun}\,\mbox{yr}^{-1}]$ & $[\mbox{km}\,\mbox{s}^{-1}]$ &
        &  $[\mbox{M}_{\sun}\,\mbox{yr}^{-1}]$ & $[\mbox{km}\,\mbox{s}^{-1}]$ \\
& & & & & & & & &\\\hline
& & & & & & & & &\\
A23 & s & $2.0\, 10^{-7}$ & 11 & s & $2.3\, 10^{-7}$ & 11 & --- &  &   \\
A25 & i & $5.0\, 10^{-6}$ & 33 & s & $5.1\, 10^{-7}$ & 18 & --- &  &    \\
A27 & i & $4.4\, 10^{-6}$ & 35 & i & $6.4\, 10^{-6}$ & 35 & i & $(6.6\, 10^{-6})$ & (40)\\
& & & & & & & & &\\
B20 & q & $9.7\, 10^{-6}$ & 16 & s & $2.0\, 10^{-7}$ & 16 & --- &  &    \\
B21 & q & $9.4\, 10^{-6}$ & 29 &i/s& $3.0\, 10^{-6}$ & 24 & i & $(1.1\, 10^{-5})$ & (28)\\
B22 & q & $1.0\, 10^{-5}$ & 34 & i & $9.6\, 10^{-6}$ & 27 & i & $(7.8\, 10^{-6})$ & (28)\\
& & & & & & & & &\\
C18 & q & $1.4\, 10^{-5}$ & 24 & i & $1.3\, 10^{-5}$ & 21 & i & $1.5\, 10^{-5}$ & 21 \\
C19 & q & $1.5\, 10^{-5}$ & 26 & i & $1.4\, 10^{-5}$ & 22 & i & $1.4\, 10^{-5}$ & 24 \\
C20 & q & $1.5\, 10^{-5}$ & 30 & i & $1.5\, 10^{-5}$ & 24 & i & $1.3\, 10^{-5}$ & 25 \\
& & & & & & & & &\\
D16 & q & $5.3\, 10^{-5}$ & 28 & i & $5.6\, 10^{-5}$ & 23 & i & $4.3\, 10^{-5}$ & 27 \\
& & & & & & & & &\\\hline
\end{tabular}
\label{tab:Resu_thre}
\end{table*}

\begin{figure*}\centering
  \includegraphics{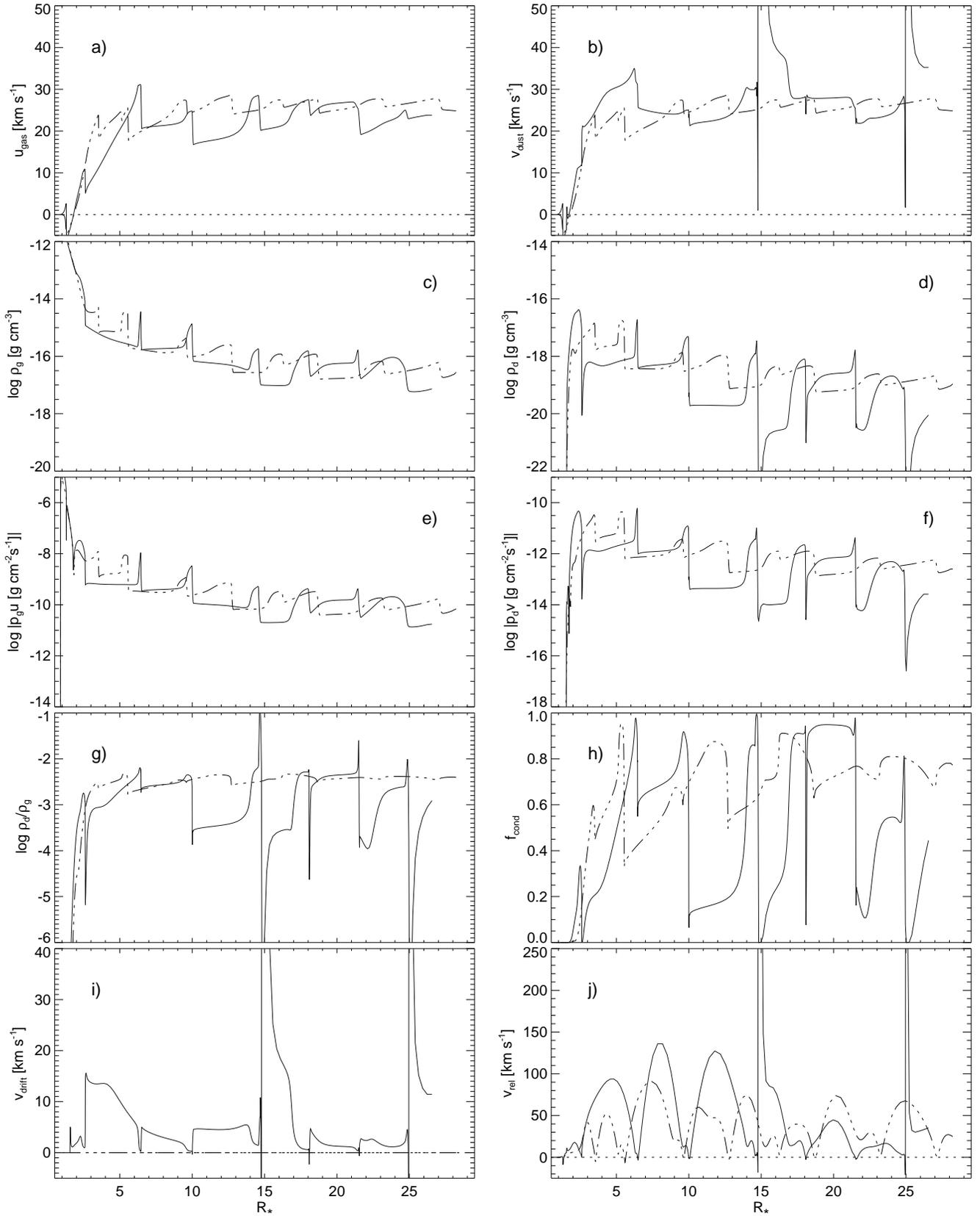}
  \caption{Spatial structures of the drift model C19d1 (solid line) and the PC model C19c2 (dash-dot-dotted line): {\bf a} gas velocity; {\bf b} dust velocity; {\bf c} gas density; {\bf d} dust density; {\bf e} gas momentum density; {\bf f} dust momentum density; {\bf g} dust-gas density ratio; {\bf h} degree of condensation; {\bf i} drift velocity; {\bf j} dust-grid relative velocity. The features in the drift model associated with the inwards facing discontinuous peaks in the dust velocity and the dust-grid relative velocity (at $15$ and $25\,R_*$) are irrelevant for the evolution of the model and should not be taken into account in physical consideration (see Sect.~\ref{sec:Nume_RHD3} and~\ref{sec:Resu_thre}).}
  \label{fig:Resu_C19cd}
\end{figure*}

We discern between local and global differences in the wind. Global differences, such as that of the mean mass loss rates will be discussed after we have looked at local differences. Figure~\ref{fig:Resu_C19cd} shows the drift model C19d1 and the PC model C19c2 at an evolved stage where the initial transient effects from the expansion phase have left the model domain. It is difficult to discuss periods of shock formation as C19d1 develops into a wind showing irregular variations, the same is true for all drift models presented here. However, typical time scales are here on the order of years. At this point we want to stress that it is the general pattern we refer to in this and all following figures of the physical structure; it is difficult to compare specific features in winds showing irregular variations. The instant presented in Fig.~\ref{fig:Resu_C19cd} has been selected because differences between the two models are seen clearly. The qualitative results that we discuss for this model are also valid for the other combinations of PC and drift models in Table~\ref{tab:Resu_thre}.

The dust density (Fig.~\ref{fig:Resu_C19cd}d) changes by about two orders of magnitude across dust fronts in C19d1, this is to be compared to the closer to one order of magnitude changes in C19c2. We can see a tendency towards narrower dust shells in C19d1 compared to C19c2 in the dust-gas density ratio, Fig.~\ref{fig:Resu_C19cd}g, and in the degree of condensation (Fig.~\ref{fig:Resu_C19cd}h) we see that regions in front of the dust fronts are sharply depleted of dust. Without the position coupling constraint dust tends to leave the low density zones and is accumulated to the regions behind shocks. A consequence of the dust relocation is that dust is shifted to regions resolved by more gridpoints. The increased gridpoint density in the regions of dust fronts allows for a numerically more accurate calculation of the advection, which is crucial when using a first order numerical scheme. The correspondence of the first order drift models in terms of more numerous and sharper shocks is better when compared to the second order PC models than they are with the first order PC models.

The dust velocity plot (Fig.~\ref{fig:Resu_C19cd}b) requires some extra explanation. As can be seen in the plot the dust has acquired enormous velocities at $15$ and $25\,R_*$. As is discussed in Sect.~\ref{sec:Nume_RHD3} these features (also seen in Fig.~\ref{fig:Resu_C19cd}f-j) appear in connection with steep dust fronts in drift models. Since they are always associated with very low dust densities they are not relevant for the model evolution and need not be taken into account in physical consideration. The cancellation of the high velocity vs\@. low density in the features is seen in the dust momentum density, Fig.~\ref{fig:Resu_C19cd}f.

The drift velocity otherwise stays fairly low. The locations of higher drift speeds are mostly the dust density troughs in the innermost region where the radiation pressure is the strongest. The maximum drift speed covering the calculated evolution of C19d1 is about $30\ \mbox{km}\,\mbox{s}^{-1}$ in inter-shock regions. Mostly, however, the upper drift speed limit is closer to $15\ \mbox{km}\,\mbox{s}^{-1}$ (as is seen in the figure). Most of the dust is found in the regions behind shocks where the drift velocity is low by the increased number of collisions with the gas. We have found the lower limit of drift speeds to be about $0.1\ \mbox{km}\,\mbox{s}^{-1}$. The speed ratio most of the time stays within the limits,
\begin{eqnarray}
-1\lesssim\log(|S_\mathrm{D}|)\lesssim1\label{eq:Resu_S}
\end{eqnarray}

%%%%%%%%%% Global effects -----------------------------------------------------
The mass loss generated in C19d1 is fairly large. In Table~\ref{tab:Resu_thre} we see that the massive winds of the C and D models are more easily formed in drift models than the less massive outflows. The evolution of the less massive winds formed in the B models are computationally more difficult. The models cover a shorter time interval and the resulting winds give less precise mean values. The parameter configurations that lead to the low mass loss time-dependent models and the stationary models in PC models (A with $\varepsilon_{\rm C}/\varepsilon_{\rm O}\le2.5$ and B with $\varepsilon_{\rm C}/\varepsilon_{\rm O}\le2.0$) do not result in a wind at all with drift included. The dust particles that initially form in the supersaturated areas of the hydrostatic atmosphere do not get a chance to grow very large. In a repeated sequence the dust first accelerates and brings the gas along. The removed position coupling together with the slow dust formation, however, prevent a further expansion. The dust is not able to support the gas, which falls back while the dust leaves the atmosphere. The condensible material in the hydrostatic initial model is not enough to drive a wind. A conclusion is that there is a higher threshold in $\varepsilon_{\rm C}/\varepsilon_{\rm O}$ when drift models form winds, compared to the PC models.

From the numbers on the mean mass loss rates and mean terminal velocities in Table~\ref{tab:Resu_thre} we see that the drift models largely reproduce the same numbers as the corresponding PC models for those parameters that lead to the formation of a wind. The local differences previously discussed therefore do not seem to affect the model globally in modifying the mean mass loss rate.\footnote{The numbers in Table~\ref{tab:Resu_thre} of the PC models do not exactly correspond to the numbers in {\rHoFeDo}, and we attribute these differences partly to the numerical differences of the codes we have used. In addition, however, we use a different dust opacity (see Sect.~\ref{sec:Nume_RHD3}) which probably accounts for most of the differences.}

The winds calculated with a first order numerical scheme do not appear periodic in any of the models we have studied. Periodicity puts very high demands on the numerical scheme. The first order scheme, maybe in combination with numerical noise allowed by the small time-steps (due to the gas shock-dust front interaction), prevents periodic models from forming by disturbing the models ``randomly''. Compare the type of wind formed in first order precise PC models with those of second order in Table~\ref{tab:Resu_thre}.

{\rSiIcDo} have found long-time modulations of several $10^2$ yrs in their model that they attribute to the dust-gas interaction. Those variations (producing structures comparable to the observed concentric arcs) should however not be confused with the variations in our model that are due to the dust induced $\kappa$-mechanism and happen on much shorter timescales of typically a few years. The authors give the stellar parameters of their model as: $M_*=0.7M_{\sun}$, $L_*=2.4\,10^4L_{\sun}$, $T_*=2010\mbox{K}$ and $\varepsilon_{\rm C}/\varepsilon_{\rm O}=1.40$. They measure a mean mass loss rate of $10^{-4}\ \mbox{M}_{\sun}\,\mbox{yr}^{-1}$ during more than nine periods, each 400 years long in their model. In one such period the star (in this case IRC +10216) would loose $0.04\ \mbox{M}_{\sun}$ and during the whole computation more than $0.36\ \mbox{M}_{\sun}$, which in turn is more than half of the stellar mass $M_*$. Therefore, the modelling of these long-term variations require a permanent ``refilling'' of the mass in the computational domain to prevent the model from changing due to the depletion of gas by the stellar wind. At present, our models do not permit such a ``refilling'' of matter, e.g\@. by transport across the inner boundary, as done by {\rSiIcDo}. Therefore the calculations have to be stopped before a significant fraction of the mass contained in the computational domain is lost (typically of a few $10^2$ years) and no studies of long-term variations can be performed at the moment, excluding a direct comparison with the result of {\rSiIcDo}. Turning to short-term variations, since we find the stellar parameters above to be unrealistic in the sense of the very low value of the effective temperature we have not made a model with these parameters.

Next, we examine the response of the model to various forms of the momentum transfer term.

%%%%%%%%%%%%%%%%%%%%%%%%%%%%%%%%%%%%%%%%%%%%%%%%%%%%%%%%%%%%%%%%%%%%%%%%%%%%%%%
%%%%%%%%%%%%%%%%%%%%%%%%%%%%%%%%%%%%%%%%%%%%%%%%%%%%%%%%%%%%%%%%%%%%%%%%%%%%%%%
%%%%%%%%%% Outcome when using different momentum transfer terms %%%%%%%%%%%%%%%
%%%%%%%%%%%%%%%%%%%%%%%%%%%%%%%%%%%%%%%%%%%%%%%%%%%%%%%%%%%%%%%%%%%%%%%%%%%%%%%
%%%%%%%%%%%%%%%%%%%%%%%%%%%%%%%%%%%%%%%%%%%%%%%%%%%%%%%%%%%%%%%%%%%%%%%%%%%%%%%
\subsection{Detailed study of the momentum transfer}\label{sec:Resu_mome}
As mentioned in Sect.~\ref{sec:Nume_RHD3} several different forms of the momentum transfer term are in use in existing stellar wind models. In all previously discussed calculations we have used the limits approximation of the drag coefficient, $C^\mathrm{LA}\ssD$, assuming fully specular collisions ($\varepsilon=1.0$). In this section we study the behavior of the C19d1 model when we on the one hand use an other approximation of the drag coefficient, and on the other hand carry out the calculations using diffusive gas-dust collisions. Table~\ref{tab:Resu_motr} gives the different model configurations.\\

\begin{table}
\caption{Detailed study of the momentum transfer term.
All models below are drift models and have the same stellar parameters as model C19d1 in Table~\ref{tab:Resu_thre}. The fraction of specular dust-gas collisions are defined by $\varepsilon$ (see Sect.~\ref{sec:Nume_RHD3}).}
\begin{tabular}{lcccc}\hline\\[-1.8ex]
model  & C$_\mathrm{D}$ (approx.)& $\varepsilon$ & $\langle\dot{M}\rangle$ & $\langle u_{\infty}\rangle$\\
&&&$[\mbox{M}_{\sun}\,\mbox{yr}^{-1}]$ & $[\mbox{km}\,\mbox{s}^{-1}]$
\\[1.0ex]\hline\\[-1.8ex]
C19d1  & $C^{\mathrm{LA}}\ssD$ & $1.0$ & $1.4\, 10^{-5}$ & 24 \\
CHV    & $C^{\mathrm{HV}}\ssD$ & $1.0$ & $1.1\, 10^{-5}$ & 22 \\
Cd0.0  & $C^{\mathrm{LA}}\ssD$ & $0.0$ & $1.3\, 10^{-5}$ & 23 \\
Cd0.5  & $C^{\mathrm{LA}}\ssD$ & $0.5$ & $1.4\, 10^{-5}$ & 23
\\[1.0ex]\hline\label{tab:Resu_motr}
\end{tabular}
\end{table}

\begin{figure}
  \resizebox{\hsize}{!}{\includegraphics{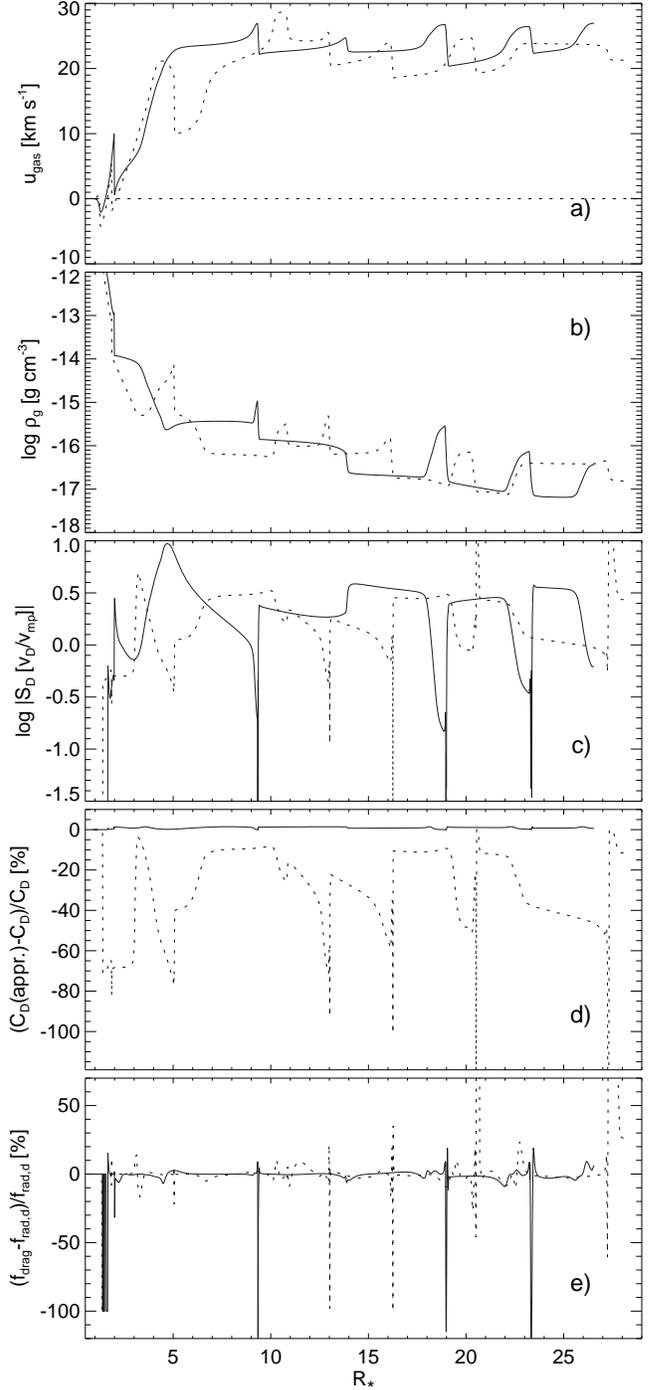}}
  \caption{Study of the high-velocity drag coefficient approximation. The models are C19d1 (solid line), and CHV (dotted line): {\bf a} gas velocity; {\bf b} gas density; {\bf c} speed ratio; {\bf d} relative deviation of the drag coefficient used in the calculations as compared to $C\ssD$; {\bf e} relative error of the structure from being CMC, the drag coefficient used in the drag force is the expression used in the calculations, i.e. $C^{\mathrm{HV}}\ssD$ for CDHV and $C^{\mathrm{LA}}\ssD$ for C19d1. The drag force is volume integrated. The features in the three lowermost panels of model C19d1 associated with the low values in the speed ratio at about $9.50$, $19.0$ and $23.5\,R_*$ should not be taken into account in physical consideration (cf\@. Fig.~\ref{fig:Resu_C19cd}). The same argument concerns model CHV. Note that the relative error of $C^{\mathrm{HV}}\ssD$ is significant in most of the envelope, especially in regions of lower speed ratios.
}\label{fig:Resu_motr}
\end{figure}

The momentum transfer term makes the RHD3-system we describe very stiff. If the system is solved explicitly there might be a reason to use an analytically simple expression for the drag coefficient such as $C^{\mathrm{HV}}\ssD$ (Eq.~\ref{eq:Nume_CDHV}). Since this approximation is in use we find it useful to compare model C19d1, which is calculated using $C^{\mathrm{LA}}\ssD$, with model CHV calculated using $C^{\mathrm{HV}}\ssD$. This is specifically motivated if we consider the measured limits of the speed ratio which we found for the models in the previous section (Eq.\ref{eq:Resu_S}). The two models are shown in Fig.~\ref{fig:Resu_motr}.

By definition the drag coefficient $C^{\mathrm{HV}}\ssD$ is not suitable in regions of low drift velocities where its use introduces a systematical error in the solution. This observation is confirmed in the diagrams of the speed ratio and the relative error of the drag coefficient (i.e\@. the drag force) used in the calculations, Fig.~\ref{fig:Resu_motr}c,d. The speed ratio $S_\mathrm{D}$ in model CHV is systematically greater in the shocked regions where that of C19d1 is low. As expected the drift velocity is the lowest in the regions behind dense shocks where the radiation pressure on the other hand is the largest. As discussed before it is in these regions where a major fraction of the dust is located.

Despite this it is not evident from the figure, and the values in Table~\ref{tab:Resu_motr}, that the global characteristics differ between CHV and C19d1. We want to point out that we cannot be certain as whether a better numerical advection or other stellar parameters will change this conclusion. In view of these results and the fact that the speed ratio does not reach high numbers we conclude that the high-velocity approximation of the drag coefficient should be used with caution in AGB wind models showing similar speed ratios.

Figure~\ref{fig:Resu_motr}e shows the relative difference of the drag force and the radiative pressure. This quantity can be used as a qualitative measure on how close CMC is achieved throughout the envelope. C19d1 is close to CMC everywhere ($\lesssim20\%$) except in the shock fronts at about $9.50$, $19.0$ and $23.5\,R_*$ where the dust velocity is lower than it would be in CMC. The same argument is valid with the shock fronts in model CHV at about $13.0$, $16.5$ and $20.5\,R_*$.

%%%%%%%%%%%%%%%%%%%%%%%%%%%%%%%%%%%%%%%%%%%%%%%%%%%%%%%%%%%%%%%%%%%%%%%%%%%%%%%
% diffusive vs. specular reflections
\begin{figure}
  \resizebox{\hsize}{!}{\includegraphics{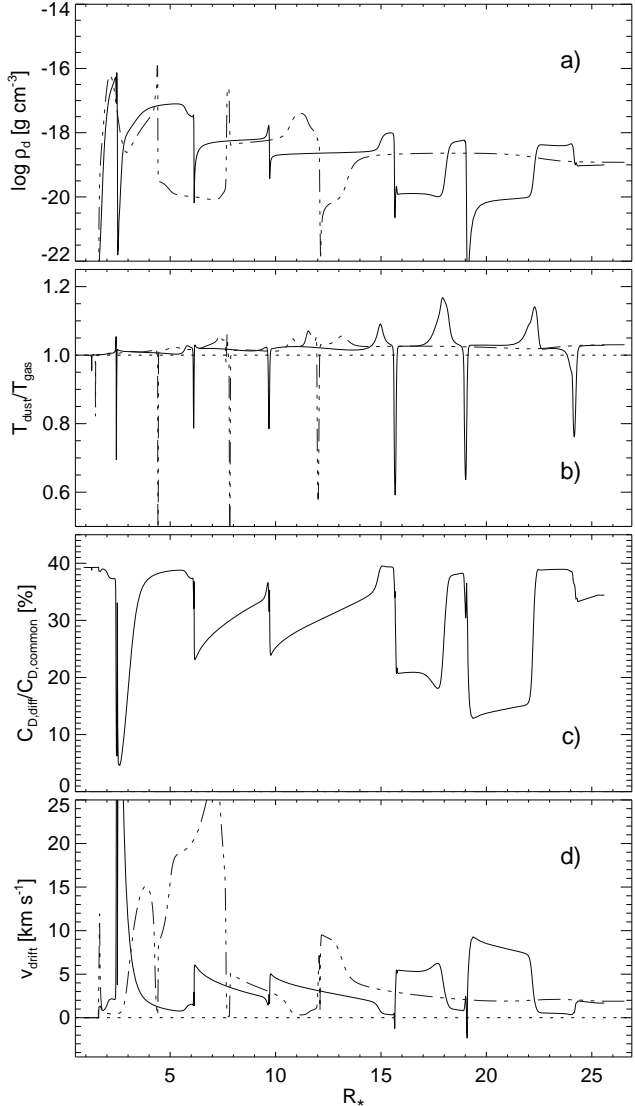}}
  \caption{Model response to two different types of collisions. 100\% diffusive collisions are assumed in model Cd0.0 (solid line). In model C19d1 (dash-dot-dotted line) they are instead 100\% specular: {\bf a} dust density; {\bf b} dust to gas temperature ratio; {\bf c} ratio of the diffusive term $C_{\mathrm{D},\mathrm{diff}}$ to the common term $C_{\mathrm{D},\mathrm{common}}$ in $C\ssD$ (Eq.~\ref{eq:Phys_CDanalytic}); {\bf d} drift velocity. The diffusive term $C_{\mathrm{D},\mathrm{diff}}$ makes up a large part of $C\ssD$ (up to $\sim40\%$). Still the overall behavior of model Cd0.0 largely agrees with model C19d1.}\label{fig:Resu_disp}
\end{figure}

We now compare models with different values on $\varepsilon$. We find that the difference in the drag coefficient between an expression describing fully diffusive and fully specular collisions under the current conditions is small. The additional term used with diffusive collisions ($C_{\mathrm{D},\mathrm{diff}}$) is proportional to the fraction of such collisions. As in the case discussed above, this term becomes important at low drift velocities in the shocks. In Fig.~\ref{fig:Resu_disp} the fully diffusive collisions model Cd0.0 (solid line) is compared with the fully specular C19d1 (dash-dotted).

The diffusive term contains the temperature ratio of the dust and the gas temperatures, Fig.~\ref{fig:Resu_disp}b. The dust temperature is equal to the radiative temperature in AGB wind models using gray opacity and assuming radiative equilibrium. Likewise the gas temperature is mostly close to the radiative temperature. In Fig.~\ref{fig:Resu_disp}c we see that the diffusive term ($C_{\mathrm{D},\mathrm{diff}}$ in Eq.~\ref{eq:Phys_CDanalytic}) is up to 40\% in size compared to the common term ($C_{\mathrm{D},\mathrm{common}}$). Again the difference is the largest in the regions behind shocks. The dust density and drift velocity plots (Fig.~\ref{fig:Resu_disp}a,d) show that the apparent effects on the structure in Cd0.0 compared to C19d1 are minute. The dust likely needs to be more accurately handled by increasing the numerical precision in the advection calculation to second order and using drift related effects in the dust formation ({\rKrSe}) before we can trace differences in dust quantities caused by the type of collisional distribution. We finally note that the differences will be even smaller with a lower fraction of diffusive collisions.

%%%%%%%%%%%%%%%%%%%%%%%%%%%%%%%%%%%%%%%%%%%%%%%%%%%%%%%%%%%%%%%%%%%%%%%%%%%%%%%
%%%%%%%%%%%%%%%%%%%%%%%%%%%%%%%%%%%%%%%%%%%%%%%%%%%%%%%%%%%%%%%%%%%%%%%%%%%%%%%
%%%%%%%%%      Conclusions
%%%%%%%%%%%%%%%%%%%%%%%%%%%%%%%%%%%%%%%%%%%%%%%%%%%%%%%%%%%%%%%%%%%%%%%%%%%%%%%
%%%%%%%%%%%%%%%%%%%%%%%%%%%%%%%%%%%%%%%%%%%%%%%%%%%%%%%%%%%%%%%%%%%%%%%%%%%%%%%
\section{Conclusions}\label{sec:Conc}
In this paper we have studied details of the dynamical interaction between dust and gas in cool stellar winds. This interaction is vital for the formation of a dust-driven outflow. In our work we have studied changes of general properties in the wind with the additional freedom of a separate equation of motion for the dust component, and we have not specifically attempted to model periodic winds.

We have here first carried out a detailed study of the physical and numerical issues associated with a separate equation of motion including mass, momentum, and energy interaction terms. The results of the drift models have been compared with position coupled (PC) models (i.e\@. models without the separate dust equation of motion).

The results of the comparisons have been divided into those of local character, i.e\@. modified spatial distribution of the dust component, and global such as time-averaged mass loss rates and the general wind behavior. Of the local results we have found that the dust is more concentrated to shocked regions in drift models; the inter-shock regions are more depleted of dust than in PC models. Globally, we have found that the mean mass loss rates of the drift models are about the same when compared with PC models for the massive winds. We have however not been able to generate the weakest winds, i.e\@. those that were stationary in the PC models. The collisional gas-dust coupling in the atmosphere of these model configurations is not strong enough to form a dust-driven stellar wind, and the dust formed instead leaves the star alone.

We have found that a high-velocity approximation of the drag coefficient in the momentum transfer term (used in wind models by e.g\@. {\rSiIcDo}) shows a systematical deviation in the drag force (compared to when using the full expression) throughout most of the envelope. The high-velocity approximation should be used with caution in winds such as those that we have modeled where the drift speed (and the speed ratio) is low in the regions where the most of the dust resides. One of the other two approximations presented, Eq.~(\ref{eq:Nume_CDSQ}) or Eq.~(\ref{eq:Nume_CDBF}), should be used instead. The winds of drift models calculated with diffusive collisions have turned out to be very similar to winds calculated with specular collisions. The interpretation should, however, not be that diffusive collisions do not make a difference. This result may turn out to be very different if frequency dependent opacities are used (instead of gray), which modify the temperature structure in the wind~\citep{Ho:99b,HoGaArJo:02,HoLoArJo:02}, and in particular the dust/gas temperature ratio (see Eq.~\ref{eq:Phys_CDanalytic}).

The three-component wind model we have presented leaves some issues open that need to be addressed before we will attempt to match observed stellar wind properties with drift models. The first issue is numerical. In order to handle the calculation of the advection more accurately, a second order advection scheme instead of the current first order is necessary; our findings on numerical convergence properties discussed in this paper (related to the two-phase shock interaction mentioned in Sect.~\ref{sec:Nume_RHD3}) could possibly also be moderated this way. Furthermore, we have so far neglected the fact that the dust formation is dependent on the drift velocity between the condensing material and the dust grains (e.g. {\rKrSe}). An improvement in the description of the formation processes is thus physically justified. A gas-dust interaction term we have not included in the current study is the heating due to friction (see Sect.~\ref{sec:Phys_inte}). We plan to discuss the importance of this term in more detail by including it in the calculations. Most AGB stars are long-period variables, and the effects of stellar pulsations should be included. This will be done in the next generation of wind models.

%%%%%%%%%%%%%%%%%%%%%%%%%%%%%%%%%%%%%%%%%%%%%%%%%%%%%%%%%%%%%%%%%%%%%%%%%%%%%%%
%%%%%%%%%%%%%%%%%%%%%%%%%%%%%%%%%%%%%%%%%%%%%%%%%%%%%%%%%%%%%%%%%%%%%%%%%%%%%%%
\begin{acknowledgements}
This work has been conducted within the framework of the school on {\it Advanced Instruments and Measurements} (AIM) supported financially by the {\it Foundation for Strategic Research} (SSF). Financial support is acknowledged from the Royal Physiographical society in Lund. All calculations have been performed on the 12-processor HPV9000 at Uppsala Astronomical Observatory, financed through a donation by the {\it Knut and Alice Wallenberg Foundation}. We thank Bengt Gustafsson and Anja C.~Andersen for carefully reading the manuscript, and the referee M.~Steffen for valuable comments and helping to clarify the origin of one of the physical terms.
\end{acknowledgements}

%%%%%%%%%%%%%%%%%%%%%%%%%%%%%%%%%%%%%%%%%%%%%%%%%%%%%%%%%%%%%%%%%%%%%%%%%%%%%%%
%%%%%%%%%% Appendix %%%%%%%%%%%%%%%%%%%%%%%%%%%%%%%%%%%%%%%%%%%%%%%%%%%%%%%%%%%
%%%%%%%%%%%%%%%%%%%%%%%%%%%%%%%%%%%%%%%%%%%%%%%%%%%%%%%%%%%%%%%%%%%%%%%%%%%%%%%
\appendix

%%%%%%%%%%%%%%%%%%%%%%%%%%%%%%%%%%%%%%%%%%%%%%%%%%%%%%%%%%%%%%%%%%%%%%%%%%%%%%%
\section{Physical conditions in the wind}\label{sec:Appe_cond}
In this subsection we comment on the justification in using the continuum approximation for the gas component in the extended atmosphere of the AGB star. We also show that the dust is in the free molecular regime when considering gas-dust collisions. Both are essential requirements in the derivation of the drag force in Sect.~\ref{sec:Phys_pit}.

\begin{figure}
  \resizebox{\hsize}{!}{\includegraphics{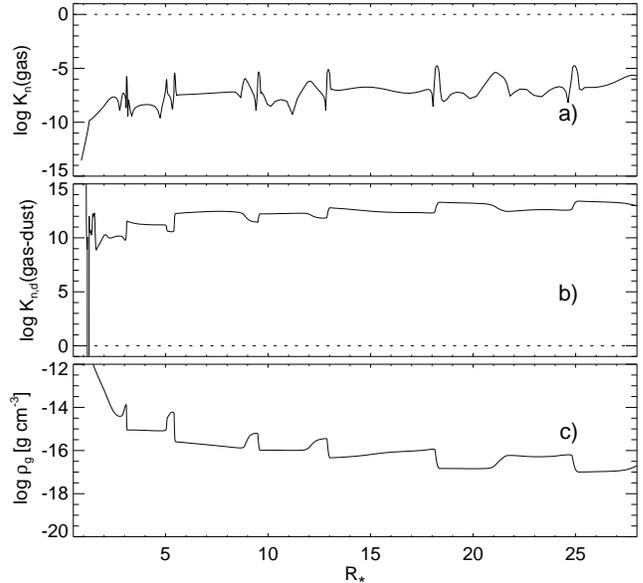}}
  \caption{Knudsen numbers of the PC model C19c2. The presented instant shows an evolved stage of the wind: {\bf a} gas Knudsen number, $\mathcal{K}_\mathrm{n}$; {\bf b} gas-dust Knudsen number, $\mathcal{K}'_\mathrm{n,d}$; {\bf c} gas density; The gas is safely in the continuum regime as $\mathcal{K}_\mathrm{n}\ll1$. The simplified gas-dust Knudsen number $\mathcal{K}'_\mathrm{n,d}$ shows that the dust is in the free-molecular regime compared to the gas. These are both essential requirements in the derivation of the expression of the drag force used here.}
  \label{fig:Appe_Kn}
\end{figure}

Both the gas and the dust are dilute in the extended atmopsheres of the AGB star. Thus,
\begin{eqnarray}
\delta\sim n^{-\frac{1}{3}}\gg d
\end{eqnarray}
where $\delta$ is the mean molecular distance; $n$ is the number density of the respective particles; $d$ is the diameter of the gas or dust particles respectively. The continuum approximation is valid for small Knudsen numbers ($\mathcal{K}_\mathrm{n}$) that satisfy,
\begin{eqnarray}
\mathcal{K}_\mathrm{n}\equiv\lambda_\mathrm{gg}/L\ll1\label{eq:Appe_Kg}
\end{eqnarray}
where $\lambda_\mathrm{gg}$ is the mean free path of a gas particle; $L$ is the characteristic size of the body or system to which the flow properties are related. A scale length of a macroscopic gradient, such as that of the density, yields a better estimate on $L$ than a constant (\rBird);
\begin{eqnarray}
L=\frac{\rho}{\mathrm{d}\rho/\mathrm{d}r}\label{eq:Appe_cond_L}\,.
\end{eqnarray}
An order of magnitude estimate of the mean free path, (assuming an elastic collision cross section $\sigma_\mathrm{g}$) is given by
\begin{eqnarray}
\lambda_\mathrm{gg}=(n\sigma_\mathrm{g})^{-1}\,.
\end{eqnarray}
Typically the collisional cross section in a gas is about $\sigma_\mathrm{g}=10^{-15}\ \mbox{cm}^2$. The Knudsen number as defined in Eq.~(\ref{eq:Appe_Kg}) is shown in Fig.~\ref{fig:Appe_Kn}a for the PC model C19c2. The Knudsen number clearly stays well below $1$ through all the extended atmosphere we model, hence we conclude that the use of the continuum approximation is justified.

The size of the dust grain ($d_\mathrm{d}$) is the body characteristic length $L$ in dust grain-gas particle collisions. The mean-free path $\lambda'_\mathrm{gd}$ of a gas particle that has (just) collided with a dust grain, that drifts through the gas ($S\ssD>0$), may be much shorter than that of the rest of the gas particles ({\rBird}). An upper limit of the associated Knudsen number assuming zero drift speed ($S\ssD=0$) is given by,
\begin{eqnarray}
\mathcal{K}_\mathrm{n,d}'=
\lambda_\mathrm{gg}/d_\mathrm{d}\label{eq:Appe_Kd}
\end{eqnarray}
$\mathcal{K}_\mathrm{n,d}'$ is shown in Fig.~\ref{fig:Appe_Kn}b, where we see that it is $\gtrsim10^{10}$ all through the atmosphere. The mean free path $\lambda'_\mathrm{gd}$ unlikely drops by a factor of ten when the speed ratio stays within the limits $0.1\lesssim S_\mathrm{D}\lesssim10$ (see Eq.~\ref{eq:Resu_S}). We conclude that the dust grain is in the free molecular flow regime compared to the gas. In the current context a free molecular flow means that the velocity distribution of a gas particle incident on a dust particle is independent of the history of previous collisions (involving other gas particles) with the same dust particle. This last statement is however not equivalent as to consider the dust by itself to be in the free molecular regime. Such a conclusion requires a measure of the dust-dust collision mean free path; a measure which we have not made in this paper.

%%%%%%%%%%%%%%%%%%%%%%%%%%%%%%%%%%%%%%%%%%%%%%%%%%%%%%%%%%%%%%%%%%%%%%%%%%%%%%%
%%%%%%%%%%%%%%%%%%%%%%%%%%%%%%%%%%%%%%%%%%%%%%%%%%%%%%%%%%%%%%%%%%%%%%%%%%%%%%%
\bibliographystyle{aa}
\bibliography{/u73/users/christer/Projects/My_Papers/CS_Refs}
\end{document}